\documentclass[preprintnumbers,eqsecnum,showpacs,aps,prd,epsf,nofootinbib,twocolumn]{revtex4}
\usepackage{latexsym,amsmath,graphicx}
\usepackage{amssymb} 
\usepackage{dcolumn}  
\newcommand{\gsim}{\mbox{\raisebox{-1.ex}{$\stackrel
      {\textstyle>}{\textstyle\sim}$}}}

\newcommand{\fourpsi}{\mbox{{$\stackrel
      {\mbox{\tiny (4)}}{\textstyle\psi}$}}}

\begin{document}
\thispagestyle{empty}
\title{Fermions on Colliding Branes}

\author{Gary Gibbons$^{1}$}
\email{G.W.Gibbons@damtp.cam.ac.uk}

\author{Kei-ichi Maeda$^{1,2}$}
\email{maeda@waseda.jp}
\author{Yu-ichi Takamizu$^{2}$}
\email{takamizu@gravity.phys.waseda.ac.jp}

\address{
\\
$^{1}$ DAMTP, Centre for Mathematical Sciences, 
University of Cambridge, 
Wilberforce Road, Cambridge CB3 OWA, UK
\\
$^{2}$ Department of Physics, Waseda University,
Okubo 3-4-1, Shinjuku, Tokyo 169-8555, Japan
}
\date{\today}

\begin{abstract}
We study the behaviour of five-dimensional
fermions localized on branes, which we 
describe by domain walls,
when two parallel branes collide in 
a five-dimensional Minkowski background
spacetime.
We find that most fermions are localized on
both branes as a whole even after collision.
However, how much fermions are localized on which brane 
depends sensitively
on the incident velocity and
the coupling constants
unless the fermions exist on both branes.
\end{abstract}

\pacs{98.80.Cq, 11.25.Wx}
\keywords{brane world, collision of branes, fermion localization}
\maketitle
\section{introduction}
It has been known since the 70's that topological defects such as
domain walls can trap fermions on their world volumes 
\cite{Jackiw_Rebbi}.
In the 80's this fact formed an integral part of suggestions
that one may regard our universe as a domain wall
\cite{Akama,Rubakov_Shaposhnikov,Visser,Gibbons_Wiltshire}, or more generally
a brane in a higher dimensional universe
\cite{Arkani,Randall_Sundrum,Binetruy_Deffayet_Langlois,Shiromizu_Maeda_Sasaki}. The idea is that the fermionic chiral matter making up  the standard model is
composed of such trapped zero modes
\cite{Gherghetta,Bajc_Gabadadze,Daemi_Shaposhnikov,Dubovsky_Rubakov_Tinyakov,Kehagias_Tamvakis,Ringeval_Peter_Uzan,Koley_Kar,Melfo_Pantoja_Tempo}. 
A similar mechanism is used in models, such as the Horawa-Witten model 
\cite{Horava_Witten,Lukas-Ovrut-Waldram} of heterotic
M-Theory, in which two
domain walls are present. Our world is localized on one brane and a
shadow world is localized on the other brane. The existence of models
with more than one brane suggests that branes may collide, and it is
natural to suppose that the Big Bang is associated with the collision
\cite{Khoury_Ovrut_Steinhardt_Turok,Steinhardt_Turok}. 
This raises the fascinating questions of what happens
to the localized fermions during such collisions?
Put more picturesquely, what is the fate of the 
standard model during brane collision? In this paper we shall embark
on what we believe is the first study of this question by solving
numerically the Dirac equation for a fermions coupled via Yukawa interaction
to a system of two colliding  domain walls, i.e. a kink-anti-kink
collision   in five-dimensional Minkowski spacetime. 
Each individual domain wall may be described analytically by a static
solution and given such a solution one may easily find analytically 
the fermion zero modes, which from the point of view of the 3+1
dimensional world volume behave like massless chiral fermions.   
The back reaction of the fermions on the domain wall is here,
and throughout this paper, neglected. 

Kink-anti-kink collisions, have
recently been  studied numerically
\cite{Anninos_Oliveira_Matzner,Takamizu_Maeda1,Copeland}. 
One solves the scalar field 
equations with initial data corresponding to a superposition of
the boosted profiles of a kink and an anti-kink. 
It was found \cite{Anninos_Oliveira_Matzner,Takamizu_Maeda1}
that, depending on the initial relative velocity that 
such domain wall pairs can pass 
through one another, or bounce, or suffer a number
of bounces in a fashion reminiscent of the cyclic universe scenario
\cite{Steinhardt_Turok}.
One may extend the treatment to include gravity 
\cite{Langlois_Maeda_Wands,Takamizu_Maeda2,Gibbons_Lu_Pope,Chen_Chong_Gibbons_Lu_Pope,McFadden_Turok_Steinhardt} 
but in this
paper we shall, for the sake of our preliminary study,
work throughout  with gravity switched off.  One may now solve the
Dirac equation in the time dependent background generated by the
kink-anti-kink collision. We use as initial data for the Dirac
equation the boosted profiles of
the chiral zero modes associated with the individual domain walls.

What we find was for us unexpected and quite remarkable.            
If the initial fermions exist on both branes, then 
without exception, for a whole range of initial
conditions, the two initially distinct but  localized fermion
distributions merge in the neighborhood of the collision, 
and then emerge after the collision again localized on  one or the other 
kink. 
By contrast if 
 one of the kinks is empty, which we refer to as
a vacuum brane, then 
 the amplitudes of the fermions
on the kinks after the collision 
highly depend on the incident velocity and the coupling constants.

\section{Fermions on moving branes}
\subsection{Fermion with Yukawa coupling and its symmetry}
\label{symmetry}
We start with a discussion of 
five-dimensional (5D) four-component fermions in a time-dependent 
domain wall in 5D Minkowski spacetime.
As a domain wall, we adopt a 5D real 
scalar field $\Phi$ with an appropriate potential $V(\Phi)$.
The 5D Dirac equation with a Yukawa coupling term 
$g \Phi\bar{\Psi}\Psi$ is given by 
\begin{align}
(\Gamma^{\hat{A}}\partial_{\hat{A}}+g \Phi)\Psi=0,~~~(\hat{A}=0,1,2,3,5)\,,
\label{Dirac_eq}
\end{align}
where $\Psi$ is a 5D four-component fermion. 
$\Gamma^{\hat{A}}$ are the Dirac matrices in 5D Minkowski spacetime 
satisfying the anticommutation relations, 
\begin{align}
\{\Gamma^{\hat{A}},\Gamma^{\hat{B}}\}=2\eta^{\hat{A}\hat{B}}\,,
\end{align}
where 
$\eta^{\hat{A}\hat{B}}={\rm diag}(-1,1,1,1,1)$ 
is Minkowski metric\footnote{
The Capital Latin indices run from 0 to 3 and 5, while 
the Greek indices from 0 to 3.}. 
We explicitly
use the following Dirac-Pauli representation
\begin{align}
&
{\Gamma}^{\hat{0}}=\left(\begin{array}{cc}
-i & 0 \\
0 & i
\end{array}
\right),~~
{\Gamma}^{\hat{5}}=\left(\begin{array}{cc}
0 & 1 \\
1 & 0
\end{array}
\right)\,,
\nonumber \\
&{\Gamma}^{\hat{k}}=\left(\begin{array}{cc}
0 & -i\sigma^k \\
i\sigma^k & 0
\end{array}
\right),~~(k=1,2,3)\,,~~
\label{Dirac_rep}
\end{align}
with $\sigma^k$ being the Pauli $2\times 2$ matrices.

Note that Eq. (\ref{Dirac_eq})
 implies current conservation law:
\begin{eqnarray}
\partial_A n^A=0\,,
\end{eqnarray}
where $n^A\equiv  \bar{\Psi}\Gamma^{\hat{A}}\Psi$ 
is conserved number current.
Here we define $\bar{\Psi}\equiv \Psi^\dagger \Gamma_{\hat{0}}$.
This gives conserved number density 
$n\equiv n^0=\bar{\Psi}\Gamma^{\hat{0}}\Psi=\Psi^\dagger \Psi$.
The total number of fermions is
defined by $N=\int d^5 X n$, which is conserved.

Later we shall need the fact that
the Dirac equation (\ref{Dirac_eq}) 
has the following 
time reversal and reflection symmetries:\\
(1)
If  $\Psi(t,\vec{x},z)$ is a solution of the Dirac equation 
with  scalar field $\Phi(t,\vec{x},z)$,
$\Gamma^{\hat{0}}\Psi(-t,\vec{x},z)$ 
is a solution of the Dirac equation with 
the scalar field $-\Phi(-t,\vec{x},z)$,
where 
$X^5=z$ is the coordinate of a fifth dimension.
 In particular, when there is no interaction ($\Phi=0$
or $g=0$),
$\Gamma^{\hat{0}}\Psi(-t,\vec{x},z)$ is time reversal of 
$\Psi(t,\vec{x},z)$
\\
(2) If  $\Psi(t,\vec{x},z)$ is a solution of the Dirac equation 
with  scalar field $\Phi(t,\vec{x},z)$,
$\Gamma^{\hat{5}}\Psi(t,\vec{x},-z)$ 
is a solution of the Dirac equation with 
the scalar field $-\Phi(t,\vec{x},-z)$.
In particular, if $\Psi(t,\vec{x},z)$ is a solution for
a kink [an anti-kink], $\Gamma^{\hat{5}}\Psi(t,\vec{x},z)$ is a solution for an
anti-kink [a kink].
It will turn out that the solution with a kink [an anti-kink]
is related to positive [negative]
chiral fermions, which are defined below (see next subsection).
\\
(3) Combining (1) and (2), we find that
$\Gamma^{\hat{5}}\Gamma^{\hat{0}}\Psi(-t,\vec{x},-z)$ 
is a solution of the Dirac equation with $\Phi(-t,\vec{x},-z)$

If we assume some symmetries for a domain wall,
we find further properties for fermions as follows.\\
(i) For the case of  a static domain wall, (1) yields
that $\Gamma^{\hat{0}}\Psi(t,\vec{x},z)$ is
a solution for an anti-kink [a kink]
if $\Psi(t,\vec{x},z)$
is a solution for a kink [an anti-kink].
\\
(ii)  If a domain wall is
described by a kink (or an 
anti-kink), which has  symmetry such that
$\Phi(t,\vec{x},-z)=-\Phi(t,\vec{x},z)$,
(2) yields
that
$\Gamma^{\hat{5}}\Psi(t,\vec{x},z)$ is
a solution for an anti-kink [a kink] if $\Psi(t,\vec{x},z)$
is a solution for a kink [an anti-kink].
\\
(iii) We may also have time symmetry such that
$\Phi(-t,\vec{x},z)=\Phi(t,\vec{x},z)$ for  collision of two walls.
In fact we find  from numerical analysis that
this ansatz is approximately correct \cite{Takamizu_Maeda1}.
Assuming $z$-reflection symmetry as well, we find from (3) that
$\Gamma^{\hat{5}}\Gamma^{\hat{0}}\Psi_\pm(-t,\vec{x},-z)$,
which is time reversal and $z$-reflection 
of $\Psi_\pm(t,\vec{x},z)$,
is also a solution for the same scalar field $\Phi (t,\vec{x},z)$.

Before going to analyze concrete examples, 
we introduce two chiral fermion states
\begin{eqnarray}
\Psi_\pm={1\over 2}\left(1\pm  {\Gamma}^{\hat{5}}\right)\Psi
\end{eqnarray}
This definition implies 
\begin{eqnarray}
{1\over 2}\left(1\pm {\Gamma}^{\hat{5}}\right)\Psi_\pm=\Psi_\pm,
~{1\over 2}\left(1\mp {\Gamma}^{\hat{5}}\right)\Psi_\pm=0\,.
\end{eqnarray}
Using the representation (\ref{Dirac_rep}), we have
\begin{align}
\Psi_+=\left(
\begin{array}{c}
\psi_+ \\
\psi_+ 
\end{array}
\right)
\,,~~
\Psi_-=\left(
\begin{array}{c}
\psi_- \\
-\psi_- 
\end{array}
\right)
\,,
\end{align}
where $\psi_+ $ and $\psi_-$ are two-component spinors.

The Dirac equation (\ref{Dirac_eq}) is now reduced to
\begin{eqnarray}
\left(\pm \partial_{\hat{5}}+g\Phi
\right)\psi_\pm
+\Gamma^{\hat{\mu}}\partial_{\hat{\mu}}
\psi_\mp
=0\,.
\end{eqnarray}


\subsection{Fermions on a kink (or an anti-kink)}
As for a domain wall, now we assume the 
potential form is given by 
 $V(\Phi)={\lambda\over 4}
\left(\Phi^2-\eta^2\right)^2$. 
Here we recall the dimension of some variables.
Since we discuss five dimensional spacetime,
we have the following dimensionality:
\begin{eqnarray}
 [\Phi]=[\eta]=L^{-3/2}, [\Psi]=L^{-2}, [g]=L^{1/2}, [\lambda]=L
\,,
\end{eqnarray}
where $L$ is a scale length.
In what follows, we use  units in which 
$m_\eta (\equiv \eta^{2/3})=1$.

Then a domain wall solution is given by
\begin{eqnarray}
\Phi=\epsilon  \tanh \left({z\over D}\right)
\label{domain_wall}
\,,
\end{eqnarray}
where $\epsilon=\pm$ correspond to a kink and an anti-kink solutions
and $D=\sqrt{2/\lambda}$ is the width of a domain wall.
Note that $\Phi(z)$ is an odd function of $z$.

As for a fermion, in the case of  a static domain wall, 
separating variables as  
$\psi_+=\fourpsi_+(x^\mu)f_+(z)$ and $\psi_-=\fourpsi_-(x^\mu)f_-(z)$
and assuming massless chiral fermions on a brane, i.e.
$\Gamma^{\hat{\mu}}\partial_{\mu}\fourpsi_\pm (x^\mu)=0$,
we find the equations for $f_\pm(z)$  as
\begin{eqnarray}
\left(\pm \partial_{5} +g\Phi(z) \right)f_\pm
=0
\end{eqnarray}

With Eq. (\ref{domain_wall}), we find the solutions are
\begin{eqnarray}
f_\pm \propto 
\left[\cosh \left({z\over D}\right)\right]^{\mp \epsilon g D}
\,.
\end{eqnarray}
Note that the fermion wave function is an even function of $z$.
Hence the positive-chiral (the negative-chiral) fermion is localized
for a kink (an anti-kink )
but is not localized for an anti-kink (a kink).

To fix numbers of fermions on a wall,
$f_\pm$ should be normalized up to 
an arbitrary phase factor $\phi_{\pm (0)}$,
which is set to be zero.
Using a number density of fermions given by
\begin{eqnarray}
n\equiv \Psi^\dagger \Psi =2\left(\psi_+^\dagger \psi_+ 
+ \psi_-^\dagger \psi_-  \right)
\,,
\end{eqnarray}
we normalize the total number of fermions localized on
a static domain wall to be unity, i.e.
$N=1$.
More precisely, for a kink (an anti-kink), we impose
\begin{align}
\int_{-\infty}^{\infty}
n_\pm d\tilde{z}=1
\,,
\end{align}
which gives
\begin{align}
f_\pm(z)=\left[{\Gamma(g D+{1\over 2})
\over 2\sqrt{\pi}D \Gamma(g D)}\right]^{1/2}
\left[\cosh \left({z\over D}\right)\right]^{- g D}
\label{sol_localization}
\,.
\end{align}

Using this solution, we can describe the wave function
of fermion localized on a kink (or an anti-kink)
as 

\begin{align}
\Psi^{(\rm K)}(x,z)
=\left(
\begin{array}{c} \fourpsi_+(x)f_+(z)
\\
\fourpsi_+(x)f_+(z)
\end{array}
\right)
\,,
\\
\Psi^{(\rm A)}(x,z)
=\left(
\begin{array}{c} \fourpsi_-(x)f_-(z)
\\
-\fourpsi_-(x)f_-(z)
\end{array}
\right)
\,.
\end{align}

To quantize the fermion fields,
we define annihilation operators
of localized fermions on a kink and on an anti-kink
by
\begin{align}
a_{\rm K}=\langle\Psi^{(\rm K)}, \Psi\rangle ~~{\rm and}~~
a_{\rm A}=\langle\Psi^{(\rm A)}, \Psi\rangle 
\end{align}
Note that those two states are orthogonal, i.e.
$\langle\Psi^{(\rm K)}, \Psi^{(\rm A)}\rangle=0$.

\subsection{Fermion wave function on a moving domain wall}

To discuss fermions at collision of branes,
we first discuss fermions on a domain wall moving with a
constant velocity.
When a domain wall is moving, however,
$\Phi$ is time-dependent, and then 
the above prescription 
(separation of the fifth coordinate) to find wave functions
is no longer valid.

Since 3-space is flat, we expand the wave functions by Fourier series
as
\begin{eqnarray}
&&
\psi_\pm  ={1\over (2\pi)^{3/2}}\int d^3\vec{k} ~e^{i \vec{k}\vec{x}}
 \psi_\pm (t, z; \vec{k})
\,.
\end{eqnarray}
We find the Dirac equations become
\begin{eqnarray}
\left(\pm \partial_{5} +g\Phi\right)
\psi_\pm 
-\left(i \partial_0 \pm (\vec{k}\cdot\vec{\sigma})
\right)\psi_\mp 
=0
\,.
\end{eqnarray}

In what follows, we shall consider only 
low energy fermions, that is,
we assume that $\vec{k} \approx 0$,
that is $|\vec{k}|$ is enough small compared
with the mass scale of 5D fermion ($g\Phi$). 
The equations we have to solve are now
\begin{eqnarray}
&&
i \partial_0 \psi_\pm =\left(
\mp \partial_{5} +g\Phi
\right)\psi_\mp
\,.
\label{eqs_psi}
\end{eqnarray}

Since up- and down-components of $\psi_\pm$ are decoupled, we discuss only 
up-components here. Note that 
taking into account $\vec{k}$  mixes the up- and down-components.
With this ansatz, we can describe fermion by two single-component 
chiral wave functions as
\begin{align}
\Psi=
\left(
\begin{array}{c} 
1\\0\\1\\0
\end{array}
\right)\psi_{+}(z,t)
+
\left(
\begin{array}{c} 
1\\0\\-1\\0
\end{array}
\right)\psi_{-}(z,t)
\,.               
\end{align} 
For a localized fermion on a static kink (or an anti-kink), 
the wave functions are
$
\psi_\pm(z,t)=f_\pm (z)
$.

Next we construct a localized fermion wave function on
 a moving domain wall with a
constant velocity $\upsilon$.
In this case, we can find the analytic solution by
a Lorentz boost.
We find for a kink with velocity $\upsilon$,
\begin{eqnarray}
\psi_+^{(\rm K)}(z,t;\upsilon)&=&\sqrt{\gamma+1\over 2}
\tilde{\psi}^{(\rm K)}
\left({\gamma(z- \upsilon t)} \right)
\label{Lorentz_fermion_kink}
\,, \\
\psi_-^{(\rm K)}(z,t;\upsilon)&=&i{\gamma \upsilon \over \gamma+1}
\sqrt{\gamma+1\over 2} 
\tilde{\psi}^{(\rm K)}
\left({\gamma(z- \upsilon t)}\right)
\nonumber
\end{eqnarray}
and for an anti-kink with velocity $\upsilon$,
\begin{eqnarray}
\psi_-^{(\rm A)}(z,t;\upsilon)&=&\sqrt{\gamma+1\over 2}
\tilde{\psi}^{(\rm A)}
\left({\gamma(z- \upsilon t)}\right)
\,,
\label{Lorentz_fermion_anti-kink}
\\
\psi_+^{(\rm A)}(z,t;\upsilon)
&=&-i{\gamma \upsilon \over \gamma+1}\sqrt{\gamma+1\over 2}
\tilde{\psi}^{(\rm A)}
\left({\gamma(z- \upsilon t)}\right)
\nonumber 
\end{eqnarray}
where $\tilde{\psi}^{(\rm K)}(\tilde{z})
=f_+ (\tilde{z})$ and $\tilde{\psi}^{(\rm A)}(\tilde{z})
=f_- (\tilde{z})$
are static wave functions of chiral fermions localized 
on static kink and anti-kink, respectively, 
and
$\gamma=1/\sqrt{1-\upsilon^2}$ is the Lorentz factor.

We can check that 
the total number of fermions is preserved
also in the boosted Lorentz frame.
From
Eqs. (\ref{Lorentz_fermion_kink}) 
and (\ref{Lorentz_fermion_anti-kink}), 
we find
that $n=\gamma \tilde{n}$.
Integrating it in the $z$-direction, we find
\begin{eqnarray}
\int_{t={\rm const}} dz ~n&=& \int_{t={\rm const}}
 dz ~\gamma \tilde{n}\left(\gamma (z-vt)
\right)
\nonumber \\
&=&\int d\tilde{z} ~\tilde{n}\left(\tilde{z}
\right)=1\,.
\end{eqnarray}

If a domain wall
is given by a kink [an anti-kink], 
we have only the positive-chiral fermions in a comoving frame
[the negative-chiral fermions]. However,
from Eqs (\ref{Lorentz_fermion_kink}) 
and (\ref{Lorentz_fermion_anti-kink}), we find that 
the negative-chiral modes [positive-chiral  modes]
also appear in this boosted Lorentz frame. 
For a kink, 
the ratio of number density of the negative-chiral modes to that of
the positive-chiral ones
is given by $\gamma^2 \upsilon^2/(\gamma+1)^2$.

The above wave functions on a moving domain wall
with constant velocity can be used for
setting the initial data for colliding domain walls.

\section{Fermions on colliding domain wall}
\subsection{Initial setup}
We construct our initial data
as follows. 
Provide a
kink solution at $z=-z_0$ and an anti-kink solution 
at $z=z_0$, which are separated by a 
large distance and approaching each other with the same speed 
$\upsilon$. 
We can set up as an initial  profile for 
the scalar field $\Phi$:
\begin{align}
\Phi(z,t)=
\Phi^{\rm (K)}(z,t;\upsilon)+\Phi^{\rm (A)}(z,t;-\upsilon)-1 \,,
\end{align}
where 
\begin{align}
\Phi^{\rm (K, A)}(z,t;\upsilon) &=\pm \tanh(\gamma (z-\upsilon t)/D) 
\end{align}
are the Lorentz boosted kink  and anti-kink solutions, respectively.
Here we have chosen that the initial time is 
$t=t_{\rm in}\equiv -z_0/\upsilon$. The domain walls 
collide at $t=0$.

For fermions on  moving walls, 
we first expand the wave function as
\begin{align}
\hat{\Psi}=\Psi_{\rm in}^{(\rm K)}(x,z;\upsilon)
a_{\rm K}
+
\Psi_{\rm in}^{(\rm A)}(x,z;-\upsilon)
a_{\rm A}
+
\Psi_{\rm in}^{(\rm B)}(x,z)
a_{\rm B}
\,,
\label{initial_state}
\end{align}
where 
$\Psi_{\rm in}^{(\rm K)}(x,z;\upsilon)$
and 
$\Psi_{\rm in}^{(\rm A)}(x,z;-\upsilon)$
are the wave function of right-moving localized fermion
on a kink and those of left-moving one on an anti-kink, respectively,
which are explicitly by 
Eq. (\ref{Lorentz_fermion_kink}) and
Eq. (\ref{Lorentz_fermion_anti-kink}).
We also denote the bulk fermions symbolically 
by $\Psi_{\rm in}^{(\rm B)}(x,z)$.
We do not give its explicit form because
it does not play any important role in the 
present situation.
We have assumed in Eq. (\ref{initial_state}) that 
$\{\Psi_{\rm in}^{(\rm K)}(x,z;\upsilon),
\Psi_{\rm in}^{(\rm A)}(x,z;-\upsilon)$ and $\Psi_{\rm in}^{(\rm B)}(x,z)\}$ 
form a complete orthogonal system.
Note that $\{\Psi_{\rm in}^{(\rm K)}(x,z;\upsilon)$ and $
\Psi_{\rm in}^{(\rm A)}(x,z;-\upsilon)\}$ are orthogonal.

Now we can set up an initial state for fermion
by creation-annihilation operators.
We shall call a domain wall associated with fermions a fermion wall,
and a domain wall in vacuum a vacuum wall.
We shall discuss
two cases: one is collision of two fermion walls, and the other is
collision of fermion and vacuum walls.
For initial state of fermions,
we consider two states;
\begin{align}
|{\rm KA}\rangle& \equiv a_{\rm A}^\dagger a_{\rm K}^\dagger|0\rangle
\,\\
|{\rm K}0\rangle& \equiv a_{\rm K}^\dagger|0\rangle
\,
\end{align}
where $|0\rangle$ is
a fermion vacuum state.

\subsection{Outgoing states and expectation values}

We discuss behaviour of fermions at collision.
After collision of two domain walls,
each wall will recede to infinity with 
almost the same velocity as the initial one $\upsilon$.
Therefore we expect that
positive chiral fermions stay 
on a left-moving kink 
 and negative ones on a right-moving anti-kink. 
Those wave functions are given by
$\Psi_{\rm out}^{(\rm K)}(x,z;-\upsilon)$
 and $\Psi_{\rm out}^{(\rm A)}(x,z;\upsilon)$.
There may be bulk fermions which 
are left behind after collision,
which wave function is symbolically written by
$\Psi_{\rm out}^{(\rm B)}(x,z)$.
Since the initial wave functions ($\Psi_{\rm in}^{(\rm K)}(x,z;\upsilon)$
and $\Psi_{\rm in}^{(\rm A)}(x,z;-\upsilon)$)
are described by the finial wave functions
($\Psi_{\rm out}^{(\rm K)}(x,z;-\upsilon)$,
$\Psi_{\rm out}^{(\rm A)}(x,z;\upsilon)$, and
$\Psi_{\rm out}^{(\rm B)}(x,z)$ at $t=t_{\rm out}\equiv z_0/\upsilon$),
we find the relations between them by solving
the Dirac equation (\ref{eqs_psi}).
Those relations can be written as 
\begin{eqnarray}
\Psi_{\rm in}^{(\rm K)}(x,z;\upsilon)
&\sim& 
\alpha_{\rm K} \Psi_{\rm out}^{(\rm K)}(x,z;-\upsilon)
+
\beta_{\rm K} \Psi_{\rm out}^{(\rm A)}(x,z;\upsilon)
\nonumber \\
&&+
\gamma_{\rm K} \Psi_{\rm out}^{(\rm B)}(x,z)
\,,
\label{Bogolubov1}\\
\Psi_{\rm in}^{(\rm A)}(x,z;-\upsilon)
&
\sim &
\alpha_{\rm A} \Psi_{\rm out}^{(\rm A)}(x,z;\upsilon)
+
\beta_{\rm A} \Psi_{\rm out}^{(\rm K)}(x,z;-\upsilon)
\nonumber \\
&&+
\gamma_{\rm A} \Psi_{\rm out}^{(\rm B)}(x,z)
\,.
\label{Bogolubov2}
\end{eqnarray}

In order to define final fermion states, 
we also describe the wave function as
\begin{align}
\hat{\Psi}=\Psi_{\rm out}^{(\rm K)}(x,z;-\upsilon)
b_{\rm K}
+
\Psi_{\rm out}^{(\rm A)}(x,z;\upsilon)
b_{\rm A}
+
\Psi_{\rm out}^{(\rm B)}(x,z)
b_{\rm B}
\label{finitial_state}\,,
\end{align}
where $b_{\rm K}$, $b_{\rm A}$
and $b_{\rm B}$
are annihilation operators of those fermion states. 
From Eqs. (\ref{initial_state}), (\ref{Bogolubov1}), (\ref{Bogolubov2}) and 
(\ref{finitial_state}), we find
\begin{align}
b_{\rm K}&=
\alpha_{\rm K} 
a_{\rm K}
+
\beta_{\rm A} 
a_{\rm A}
\,,
\label{Bogolubov3}\\
b_{\rm A}&=
\alpha_{\rm A} 
a_{\rm A}
+
\beta_{\rm K} 
a_{\rm K}
\label{Bogolubov4}
\end{align}

Using the Bogoliubov coefficients
$\alpha_{\rm K},  \beta_{\rm K}$ and
$\alpha_{\rm A}, \beta_{\rm A} $, we obtain
the expectation values of 
fermion number on a kink and an anti-kink
after collision as
\begin{align}
\langle N_{\rm K}
\rangle
\equiv \langle{\rm KA}|b_{\rm K}^\dagger b_{\rm K}|{\rm KA}\rangle=
|\alpha_{\rm K}|^2+|\beta_{\rm A}|^2
\,\\
\langle N_{\rm A}\rangle
\equiv \langle{\rm KA}|b_{\rm A}^\dagger b_{\rm A}|{\rm KA}\rangle=
|\alpha_{\rm A}|^2+|\beta_{\rm K}|^2
\end{align}
for the case of $|{\rm KA}\rangle$.
If the initial state is $|{\rm K}0\rangle$,
we find
\begin{align}
\langle N_{\rm K}\rangle
&\equiv \langle{\rm K}0|b_{\rm K}^\dagger b_{\rm K}|{\rm K}0\rangle=
|\alpha_{\rm K}|^2
\,\\
\langle N_{\rm A}\rangle 
&\equiv \langle{\rm K}0|b_{\rm A}^\dagger b_{\rm A}|{\rm K}0\rangle=
|\beta_{\rm K}|^2
\,.
\end{align}

\subsection{Time evolution of fermion wave functions}
In order to obtain the Bogoliubov coefficients,
we have to solve 
the equations for domain wall $\Phi$\cite{Takamizu_Maeda1} and 
fermion $\Psi$ numerically.
 For the time evolution of $\Psi$, 
we use the Crank-Nicholson method since it 
is generally shown to be useful for the parabolic 
type of partial differential equation. 

In our
simulation of  two-wall collision,
we have three unfixed parameters, i.e. 
a wall thickness ($D$) 
and an initial wall velocity (${\upsilon}$) and 
a coupling between fermions and a domain wall ($g$). 
From the solution (\ref{sol_localization}), 
we find the fermions are localized within the domain wall width 
$D$ if $g ~\gsim ~2/D$.
When $g < 2/D$, fermions leak out from the domain wall.
Hence, in this paper, we analyze for the case of $g\geq 2$.
We set $D=1$, but leave $\upsilon$ free.

Before showing our results for fermions, we
summarize the behaviours of domain walls discussed in 
\cite{Takamizu_Maeda1}.
We find a bounce or a few bounces at the collision of domain walls,
which depends in a complicated way  on the initial velocity 
(There is a fractal structure in the initial velocity space
\cite{Anninos_Oliveira_Matzner}).
After the collision, two domain walls recede into infinity
with almost same velocity $\pm \upsilon $.
It is similar to collision of solitons.

To obtain the Bogoliubov coefficients,
we solve the Dirac equation 
for the  collision of fermion-vacuum walls, i.e.
fermions are initially localized on one wall,
and the other wall is empty 
($\Psi_{\rm in}^{(\rm K)}(x,z;\upsilon)$
 or $\Psi_{\rm in}^{(\rm A)}(x,z;-\upsilon)$ ).

We shall give numerical results only for the case that
positive chiral fermions are initially localized on a kink
($\Psi_{\rm in}^{(\rm K)}(x,z;\upsilon)$).
Because of $z$-reflection symmetry discussed in \S.\ref{symmetry},
we find the same Bogoliubov coefficients
for the case that
negative chiral fermions are initially localized on an
anti-kink
($\Psi_{\rm in}^{(\rm A)}(x,z;-\upsilon)$), i.e. 
$|\alpha_{\rm K}|^2=|\alpha_{\rm A}|^2$ and
$|\beta_{\rm K}|^2=|\beta_{\rm A}|^2$.

\begin{figure}[h]
\begin{center}
\begin{tabular}{c}
 \scalebox{0.15}{\includegraphics[angle=-90]{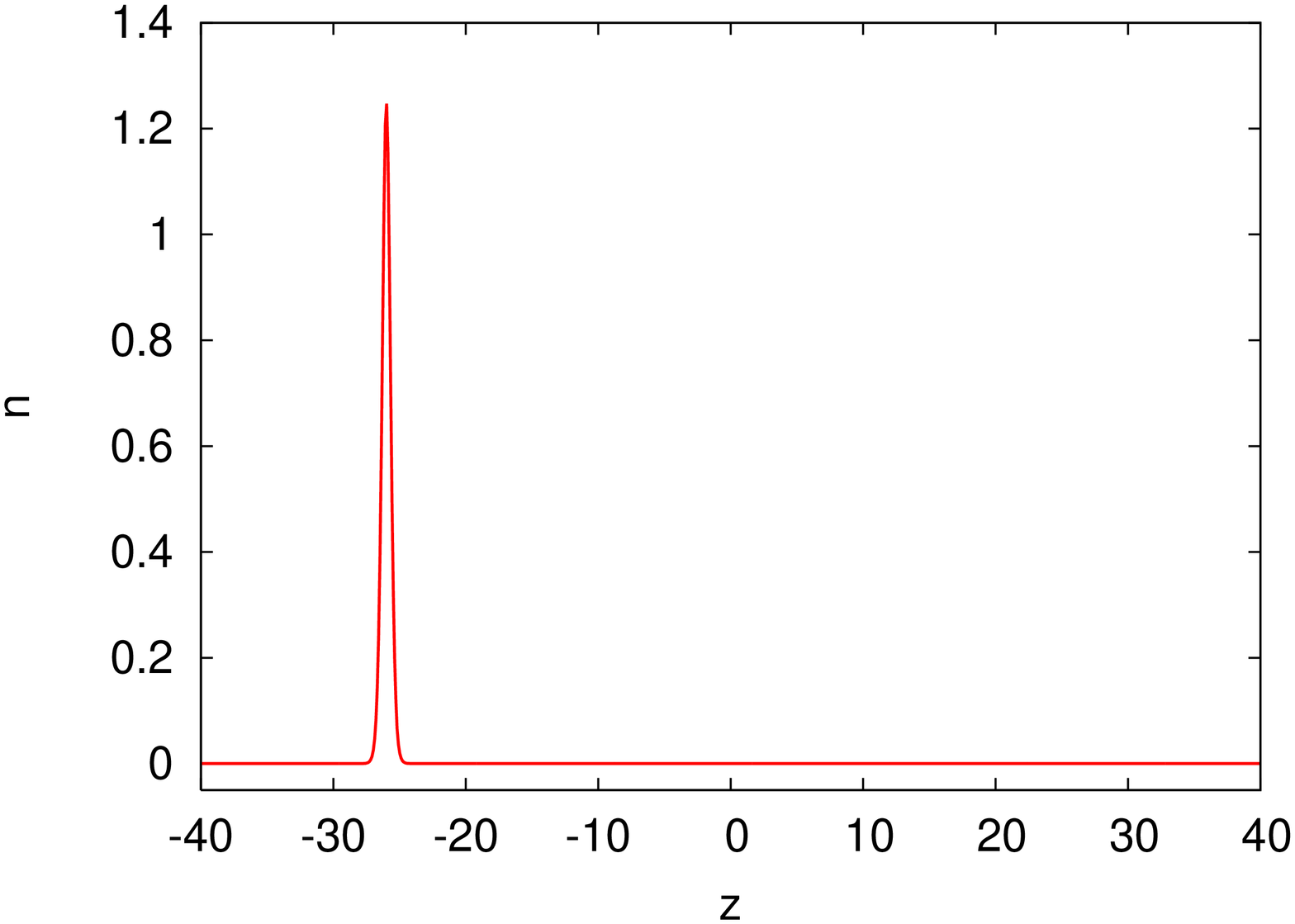}}
 \scalebox{0.15}{\includegraphics[angle=-90]{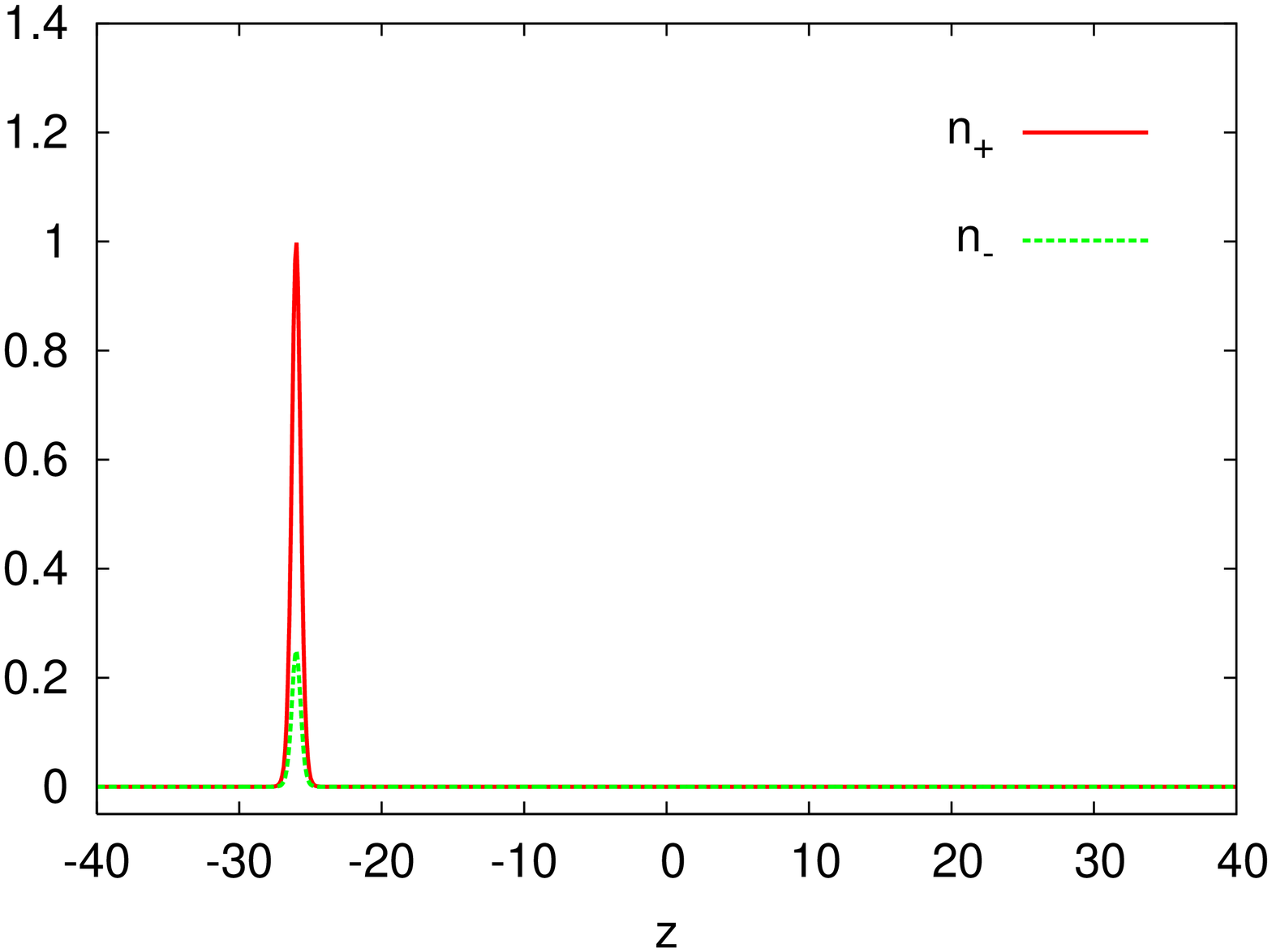}}
\\ 
(a) $t=0$ (initial)
\\
\scalebox{0.15}{\includegraphics[angle=-90]{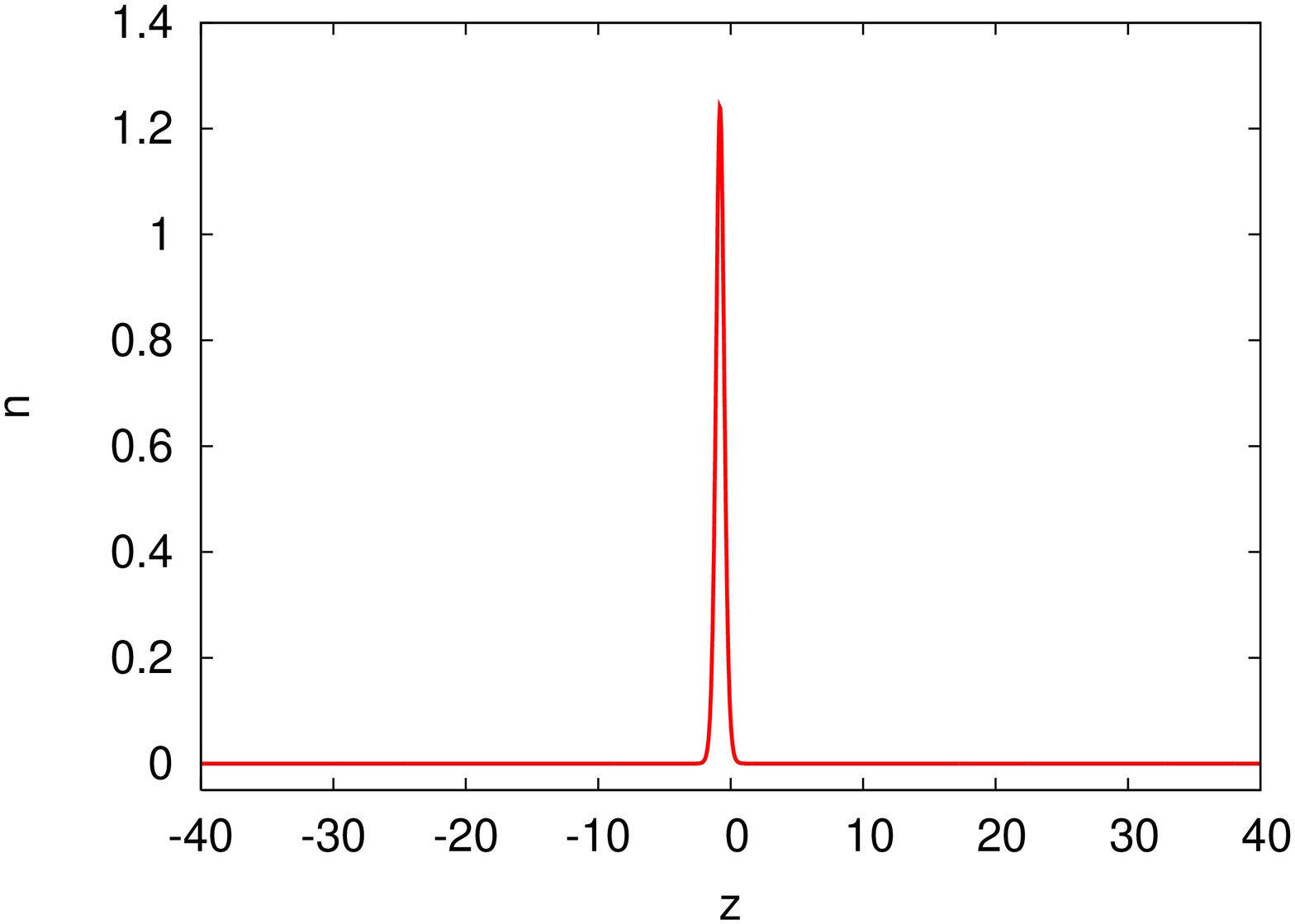}}
\scalebox{0.15}{\includegraphics[angle=-90]{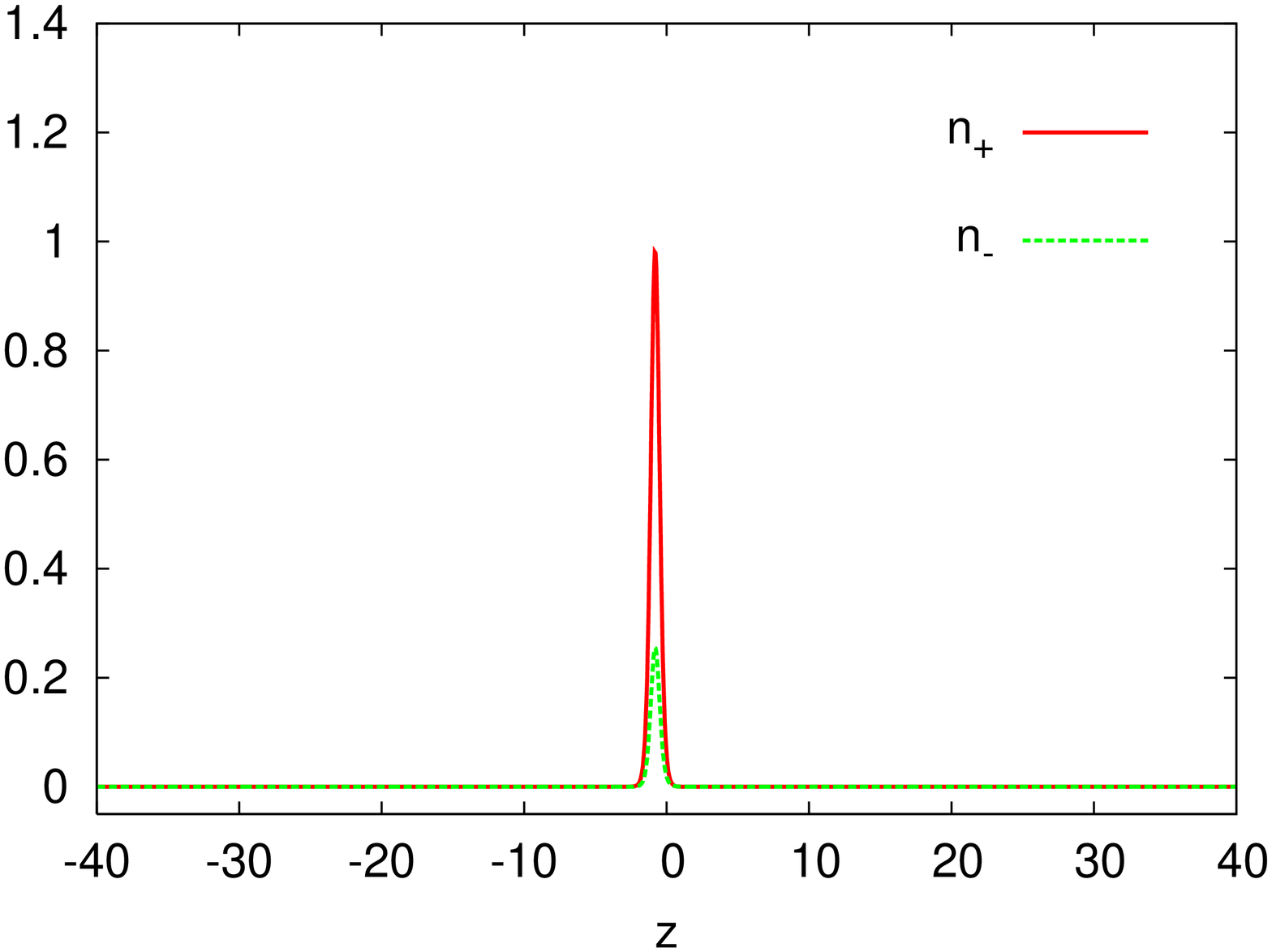}}
\\ 
(b) $t=32$ (at collision)
\\
 \scalebox{0.15}{\includegraphics[angle=-90]{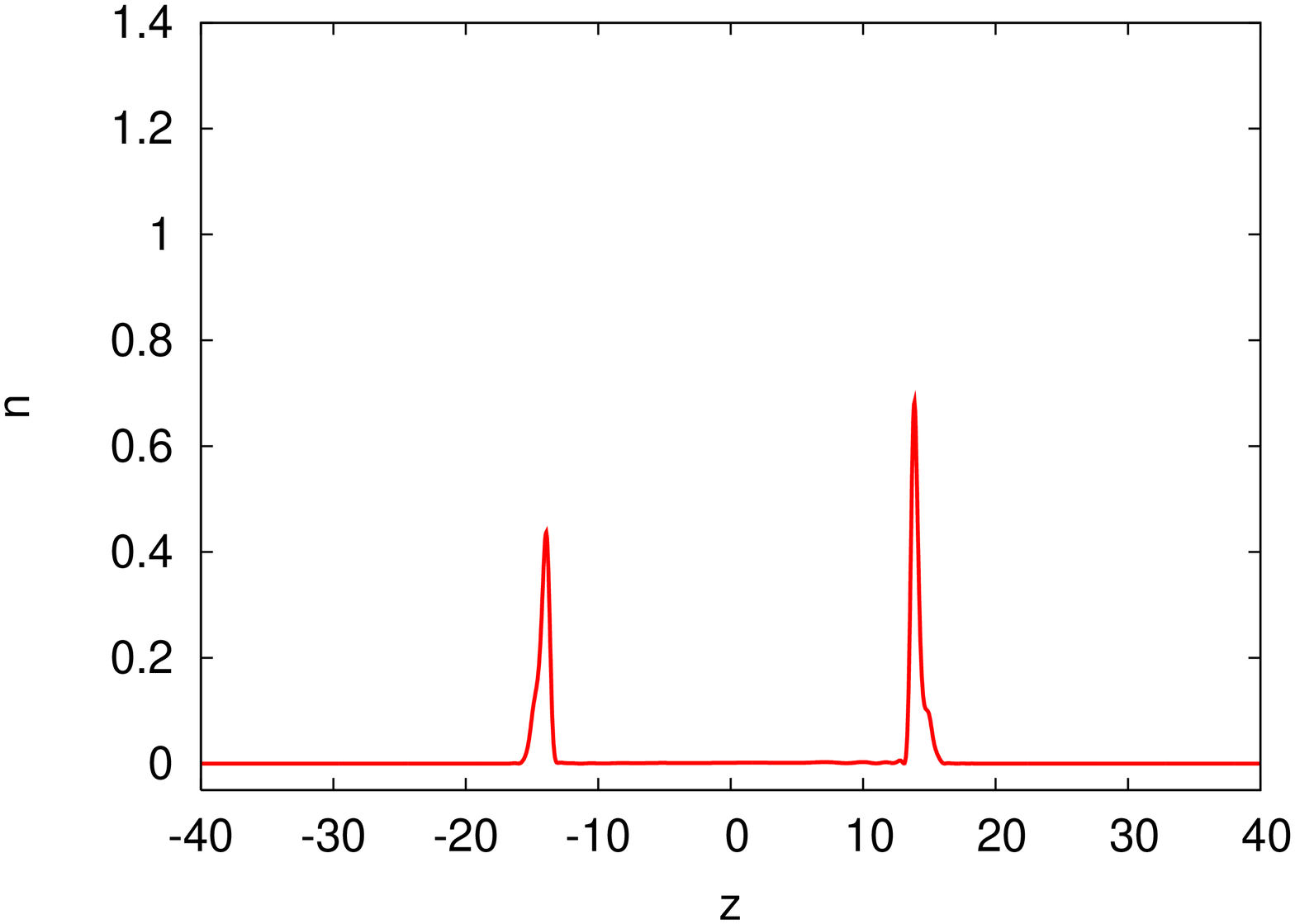}}
 \scalebox{0.15}{\includegraphics[angle=-90]{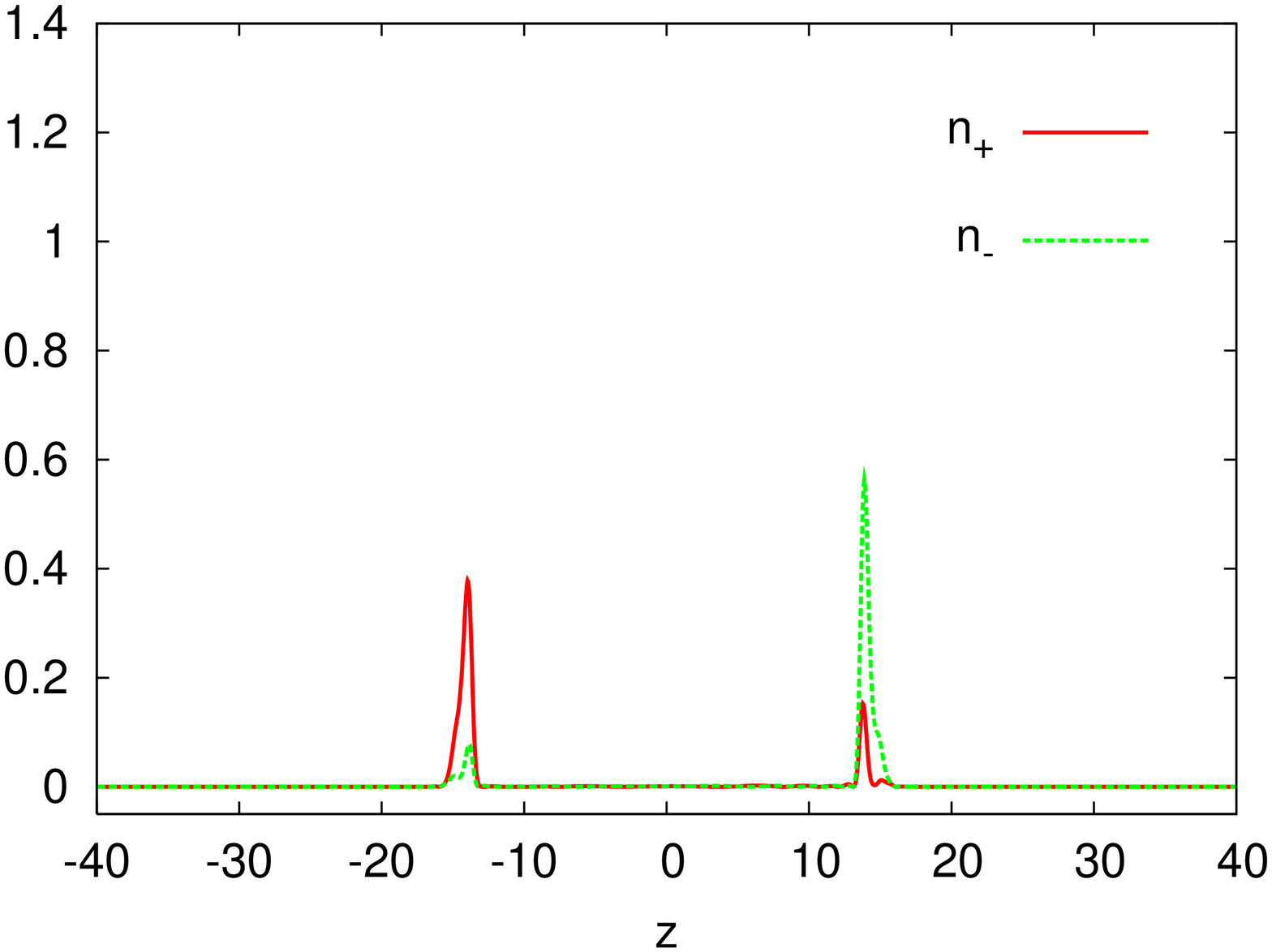}}
\\
 (c) $t=56$ (after collision)
\\
\end{tabular}
 \caption[fig1]{
Snap shots of the number density of the wave function
($n=\Psi^\dagger \Psi$) and those of
two chiral states
$n_\pm$ 
for collision of fermion-vacuum walls.
We set 
$D=1,g=2$, and $\upsilon=0.8$.}
\label{fig1}
\end{center}
\end{figure}

Setting $g=2$ and $\upsilon=0.8$, 
we show the result in Fig. \ref{fig1}. 
The other chiral mode appears at collision
and the  wave function 
splits into two parts after collision.

 From the asymptotic behaviour of the wave function
as $t\rightarrow \infty$, we
obtain the Bogoliubov coefficients 
numerically such that $|\alpha_{\rm K}|^2
=0.42$ and $|\beta_{\rm K}|^2=0.55$.
Since a few amount of fermions escapes into bulk space at collision,
$|\alpha_{\rm K}|^2+|\beta_{\rm K}|^2$ is not conserved, and 
the difference between the initial value and the final one 
($|\gamma_{\rm K}|^2=1-(|\alpha_{\rm K}|^2+|\beta_{\rm K}|^2)$)
corresponds to the amount of bulk fermions left behind.

The Bogoliubov coefficients depend on the initial wall 
velocity.
In Table \ref{table1}, we summarize our results for
different values of velocity.

\begin{table}[h]
	\begin{center}
		\begin{tabular}{|c|c c c |c cc |}
		\hline
		& 
		&$g=2 $
		& 
		& 
		&$g=2.5$ 
		&
		  \\
\cline{2-7}
 \raisebox{1.5ex}{$\upsilon$} 
	 	&  $|\alpha_{\rm K}|^2$
		& $|\beta_{\rm K}|^2$ &$|\gamma_{\rm K}|^2$
		 & $|\alpha_{\rm K}|^2$
                &$|\beta_{\rm K}|^2$ &$|\gamma_{\rm K}|^2$\\
		\hline
		 {0.3} 
		& 0.94
		& 0.056
		& 0.004
		& 0.47
		& 0.53
		& 0.00
		\\
		\cline{1-7}
	 	 {0.4} 
		& 0.87
                & 0.12
		& 0.01
		& 0.57
		& 0.40
		& 0.03
		\\
		\cline{1-7}
		 {0.6} 
		& 0.69
		& 0.30
		& 0.01
		& 0.78
		& 0.17
  		& 0.05
		\\
		\cline{1-7}
	 	 {0.8} 
		& 0.42
		& 0.55
		& 0.03
		& 0.88
		& 0.02
		& 0.10
		\\
		\hline
		\end{tabular}
	\end{center}
	\caption{The Bogoliubov coefficients
of fermion wave functions localized on 
each domain wall after collision 
( $|\alpha_{\rm K}|^2$ and  $|\beta_{\rm K}|^2$)
 with
respect to the initial velocity  $\upsilon$.
We also show the amount of fermions escaped into bulk space
($|\gamma_{\rm K}|^2=1-(|\alpha_{\rm K}|^2+|\beta_{\rm K}|^2)$).}
	\label{table1}
\end{table}

\begin{figure}[h] 
\begin{center}
\begin{tabular}{c}
 \scalebox{0.36}{\includegraphics[angle=-90]{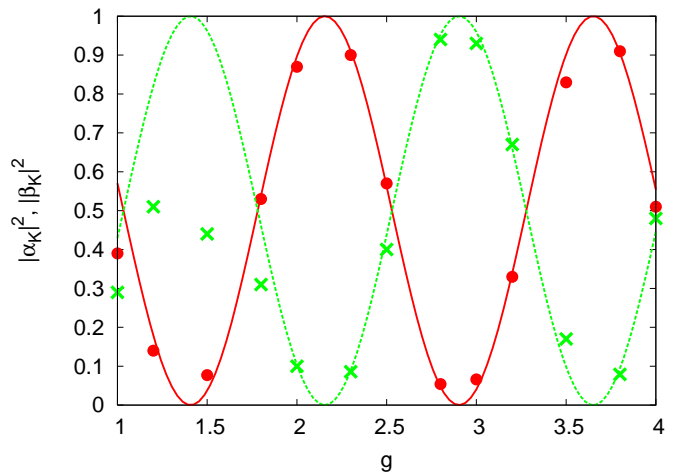}}
\end{tabular}
 \caption[fig2]{
The Bogoliubov coefficients ($|\alpha_{\rm K}|^2,
|\beta_{\rm K}|^2$)
 with 
$\upsilon=0.4$ in terms of a coupling constant $g$.
The circle and the cross denote $|\alpha_{\rm K}|^2$
and
$|\beta_{\rm K}|^2$ respectively.
Two sine curves 
($|\alpha_{\rm K}|^2,
|\beta_{\rm K}|^2 \approx [1\pm \sin (4.2 g -1.2)$)]/2
show the formula (\ref{sol_A})
with the best-fit parameters. 
} \label{g-dep}
\end{center}
\end{figure}

We also show the case of $g=2.5$ in Table \ref{table1}.
For the coupling constant $g=2$, 
$|\alpha_{\rm K}|^2$ and  $|\beta_{\rm K}|^2$
are almost equal
($0.44$ and $0.55$),
but for  $g=2.5$, most fermions 
remain on the kink ($|\alpha_{\rm K}|^2=0.88$
 and $|\beta_{\rm K}|^2=0.02$).
We find that
the Bogoliubov coefficients depend sensitively
on the coupling constant $g$ as well as the
velocity $\upsilon$.
In Fig. \ref{g-dep}, we shows the $g$-dependence.

Since the wave function is changed at collision,
when the background scalar field evolves
in a complicated way,
one might think that the behaviour of wave function
would be
difficult to describe analytically.
However, we may understand the qualitative 
behaviour in terms of  the following naive discussion.

Before collision, the wave function is approximated
well
by $\Psi_{\rm in}^{(\rm K)}(x,z;\upsilon)$.
In order to evaluate the wave function of fermion
after collision,  we have to integrate the Dirac equation (\ref{eqs_psi}).
During the collision, the spatial distributions of fermion
wave functions are 
well-described by some symmetric function
of the $z$-coordinate (see Fig. \ref{fig1} (b)).
 So we may approximate them as
\begin{eqnarray}
\psi_{\pm}=A_\pm(t) e^{i\phi_{\pm}(t)}
\psi_{0}(z)
\,,
\end{eqnarray}
where $\psi_{0}(z)$ is a normalized even real function.
$A_\pm$ and $\phi_\pm$ are regarded as the amplitudes of positive-
(negative-) chiral modes and those phases, respectively.
The scalar field $\Phi$ evolves as $\Phi : 1 \rightarrow \Phi_c (\approx 
-1.5) \rightarrow 1$ at the collision point ($z=0$).
If we approximate the scalar field as $\Phi=\Phi_c$ at collision
for collision time $\Delta t (\sim D /c)$,
integration of Eq.(\ref{eqs_psi}) with respect to $z$ gives
the change of amplitudes and phases of wave functions
as
\begin{eqnarray}
&&{1\over \sqrt{1-A_\pm^2}}\partial_0 A_\pm =\pm g\Phi_c \sin 
(\Delta \phi)\,,
\label{eq_A}
\\
&&\partial_0\phi_\pm=-{g\Phi_c}{\sqrt{1-A_\pm^2}\over A_\pm}\cos
(\Delta \phi)\,,
\label{eq_phi}
\end{eqnarray}
where $\Delta \phi\equiv \phi_- -\phi_+$.
We have also assumed
that total amplitude of wave functions is normalized ($A_+^2 +A_-^2=1$).
This means that we ignore bulk fermions, which 
may be justified because $|\gamma_{\rm K}|^2\ll 1$.
If $\Delta \phi=0$ and $A_\pm=1$
initially, then we find $A_+ (\Delta t)=1$
(or $A_- (\Delta t)=1$)
from Eq. (\ref{eq_A}), which guarantees
$\Delta \phi=0$ anytime from Eq. (\ref{eq_phi}).

We find that 
$(A_+, \Delta \phi)=(1,0)$
(or $(A_-, \Delta \phi)=(1,0)$) is a fixed point
of the system (Eqs. (\ref{eq_A}) and (\ref{eq_phi})).
However, it turns out that those are unstable.
On the other hand, we find that 
$\Delta \phi = \pi/2$  (or $-\pi/2$)
is an attractor (stable fixed points)
of the present system.
The time scale to approach these attractors 
is given by 
$(g|\Phi_c|)^{-1}$ if $\Delta A^2\equiv
A_-^2-A_+^2=O(1)$.

Once we assume  $\Delta \phi=\pm \pi/2$, then 
we find that the phases $\phi_\pm$ do not change.
Then 
we can integrate 
Eq. (\ref{eq_A}), finding
\begin{align}
A^2_{\pm }(\Delta t)={1\over 2}
\left[
1\pm \sin\left(
2\varepsilon g\Phi_c \Delta t 
+C_0
\right)
\right]
\,,
\label{sol_A}
\end{align}
where $\varepsilon=\pm 1$ and $C_0$ is an integration constant.

This formula may provide a rough evaluation 
of $|\alpha_{\rm K}|^2,
|\beta_{\rm K}|^2$.
Comparing the numerical data and 
the formula (\ref{sol_A}) with $\Phi_c\approx -1.5$, we find  
the fitting  curves in Fig. \ref{g-dep}
($\varepsilon=-1$,
$\Delta t\approx 1.4$ and $C_0=-1.2$).
The above naive analysis explains  our results 
very well.
We then conclude that  
$\Delta\phi=\pm \pi/2$ is generic except for
a highly symmetric and fine-tuned initial setting
($A_+=1$ or $A_-=1$ and $\Delta\phi=0$),
and the formula (\ref{sol_A}) 
with $\Delta\phi=\pm \pi/2$ is eventually found 
after collision.
The small difference may be understood 
by the details of the complicated 
dynamics of colliding walls.

\subsection{Fermion numbers on domain walls after collision}

We can evaluate the expectation values of
fermion numbers after collision as follows.
For the  initial state of fermions,
we consider two cases: case (a) collision of two fermion walls
$|{\rm KA}\rangle$
and case (b)
collision of fermion and vacuum walls
$|{\rm K}0\rangle$.

In the case (a), we find
\begin{align}
\langle N_{\rm K}\rangle &=
|\alpha_{\rm K}|^2+|\beta_{\rm A}|^2
=|\alpha_{\rm K}|^2+|\beta_{\rm K}|^2
\nonumber \\
&=1-|\gamma_{\rm K}|^2
\approx 1
\,\\
\langle N_{\rm A}\rangle &=
|\alpha_{\rm A}|^2+|\beta_{\rm K}|^2
=|\alpha_{\rm A}|^2+|\beta_{\rm A}|^2
\nonumber \\
&=1-|\gamma_{\rm A}|^2
\approx 1
\,.
\end{align}
We find that most fermions on domain walls
remain on both walls even after the collision.
A small amount of fermions escapes into the bulk spacetime
at collision.

In the case (b), however,
we obtain
\begin{align}
\langle N_{\rm K} \rangle =
|\alpha_{\rm K}|^2
~\,, ~~
\langle N_{\rm A}\rangle =
|\beta_{\rm K}|^2
\,.
\end{align}
Since the Bogoliubov coefficients
depend sensitively on both  the velocity $\upsilon$
and the coupling constant $g$,
the amount of fermions on each wall
is determined by the fundamental model
as well as
the details  of the  collision of the  domain walls.

\section{Concluding Remarks}
We have studied  the behaviour of five-dimensional
fermions localized on  domain walls,
when two parallel walls collide in five-dimensional Minkowski background
spacetime.
We have analyzed the dynamical behavior of fermions during collision of 
fermion-fermion branes (case (a)) and that of fermion-vacuum ones (case (b)). 

In order to evaluate expectation values of fermion number on a kink and an 
antikink after collision, 
we solve the Dirac equation for the wave function in the case (b)
and find the  Bogoliubov coefficients, in which $\beta_{\rm K}$ 
denotes the amount of fermions transfering from a kink to an 
antikink (a vacuum wall).
As a result, in the case (b) some fermions jump up to the vacuum brane at 
collision. The amount of fermions localized on which brane depends sensively 
on the incident velocity and the coupling constants $g/\lambda$ where $g$ and 
$\lambda$ are the Yukawa coupling constant and that of the 
double-well potential, 
respectively. It can be intuitively understood that the amount of 
localized fermions 
is roughly determined by the duration of collision for which they 
transfer to another wall or stay on the initial wall, and the localization 
condition depending on the 
Yukawa coupling constant between fermions and domain walls.

On the other hands, in the case (a), 
we find that most fermions seem to stay on 
both branes even after collision. This is because of  the relationship 
$|\beta_{\rm K}|=|\beta_{\rm A}|$, which is guaranteed 
by a left-right symmetry in the present system.
 This result means 
physically that the same state ($k\sim 0$) of fermions are exchanged for each 
other by the same amount. Therefore, the final amounts of fermions
 does not depend on parameters.

We conclude with some  comments about the subject 
not mentioned  above:\\[.5em]
(1) For the case of $g<2/D$, the localization of fermions on 
a domain wall is not sufficient. The tail of fermion distribution
extends  outside the wall.  As a result, we find that a  
considerable amount of  fermions escapes into a bulk space at collision.
For example, we find $|\alpha_{\rm K}|^2+|\beta_{\rm K}|^2
=0.64$ for $g=1$ and $\upsilon=0.8$. 
The formula (\ref{sol_A}) is also no longer valid in this case
(see Fig. \ref{g-dep}).
This is because localization is not sufficient.
\\[.3em]
(2) 
The collision of domain walls is rather complicated.
We find a few bounces at collision depending on 
the incident velocity. The number of bounces
is determined in a complicated way (a fractal structure
in the initial phase space \cite{Anninos_Oliveira_Matzner,Takamizu_Maeda1}).
Thus  if we change the incident velocity very little,
the number of bounces changes.
This causes a drastic change of final distribution of
fermions on each wall for the case (b).
\\[.3em]
(3) 
Since we have discussed  only the case of zero-momentum 
fermion on branes ($\vec{k}=0$),
we have only a single state on each brane, which 
constrains the fermion number to be less than unity.
If we take into account degree of freedom of low energy fermions,
we can put different states of fermions on each brane.
As the result, 
the final state of fermions after collision 
is different from the initial state, 
and it depends sensitively on
the coupling constant as well as the initial wall 
velocity just as 
the case of collision of fermion-vacuum walls.
\\[.3em]
(4) 
In the case of collision of two vacuum branes, 
nothing happens in the present approximation. 
The pair production of fermion 
and antifermion, for which we have to take into account
the momentum $k$, may occur at collision. 
This pair production process may  aslo be 
important in the cases of collision of two fermion branes
and that of fermion-vacuum branes. 
The work is in process.
\\[.3em]
(5) 
Inclusion of self-gravity is important.
It changes the fate of domain wall collision \cite{Takamizu_Maeda2}.
It would be  interesting to see what happens to the  fermion distribution
when we have a  singularity.
Although we are now analyzing it based on a supergravity model \cite{eto_sakai},
it may be more important to study the model based
on superstring or M-theory.

We will publish the results elsewhere.

\acknowledgments

We would like to thank Valery Rubakov
for valuable  comments and discussions.
This work was partially supported by the Grant-in-Aid for Scientific Research
Fund of the JSPS (No. 17540268 and 17-53192) and for the
Japan-U.K. Research Cooperative Program,
and by the Waseda University Grants for Special Research Projects and
 for The 21st Century
COE Program (Holistic Research and Education Center for Physics
Self-organization Systems) at Waseda University.
KM would like to thank DAMTP and the Centre for Theoretical Cosmology
for hospitality during this work and Trinity College for a Visiting Fellowship.
We would also acknowledge hospitality of the Institute of Cosmology 
and Gravitation, University of Portsmouth, where this work was completed,
supported by PPARC visiting grant PP/D002141/1.



\begin{thebibliography}{99}
\bibitem{Jackiw_Rebbi} 
	R. Jackiw and C. Rebbi, Phys. Rev. D, {\bf 13} (1976) 3398.
\bibitem{Akama}
	 K. Akama,
	 Lect. Notes Phys. {\bf 176}, 267 (1982).
\bibitem{Rubakov_Shaposhnikov}
	 V. A. Rubakov and M. E. Shaposhnikov,
	 Phys. Lett. {\bf B152}, 136 (1983).
\bibitem{Visser} M. Visser, Phys. Lett. {\bf B159}, 22 (1985)
	[hep-th/9910093].
\bibitem{Gibbons_Wiltshire}
	G.W. Gibbons, D.L. Wiltshire,
	Nucl. Phys. {\bf B287}, 717 (1987) [hep-th/0109093]
\bibitem{Arkani}
	 N. Arkani-Hamed, S. Dimopoulos, and G. Dvali, 
	 Phys. Lett.  {\bf B429}, 263 (1998) [hep-ph/9803315];\\
	 I. Antoniadis, N. Arkani-Hamed, S. Dimopoulos, and G. Dvali,
	 Phys. Lett. {\bf B436},  257 (1998) [hep-ph/9804398 ];\\
	 N. Arkani-Hamed, S. Dimopoulos and G. Dvali, Phys. Rev. D {\bf 59},
         086004	(1999) [hep-ph/9807344];\\
	 N. Arkani-Hamed, S. Dimopoulos, N. Kaloper, and J. March-Russell,
	 Nucl. Phys. {\bf B567},189 (2000)  [hep-ph/9903224].
\bibitem{Randall_Sundrum}
	 L. Randall and R. Sundrum,
	 Phys. Rev. Lett. {\bf 83}, 4690 (1999) [hep-th/9906064];\\
	 L. Randall and R. Sundrum,
	 Phys. Rev. Lett. {\bf 83}, 3370 (1999) [hep-ph/9905221].
 \bibitem{Binetruy_Deffayet_Langlois}
	 P. Bin\'etruy,  C. Deffayet, and Langlois,
	 Nucl. Phys.{\bf 565}, 269 (2000) [hep-th/9905012];\\
	P. Bin\'etruy,  C. Deffayet, U. Ellwanger, D. and Langlois,
	Phys. Lett. {\bf B477}, 285 (2000) [hep-th/9910219].
 \bibitem{Shiromizu_Maeda_Sasaki}
	 T. Shiromizu, K. Maeda, and M. Sasaki,
	 Phys. Rev. D {\bf 62}, 024012 (2000) [gr-qc/9910076].
\bibitem{Gherghetta}
	T. Gherghetta and A. Pomarol, Nucl. Phys. {\bf B586} (2000) 141
	[hep-ph/0003129].
\bibitem{Bajc_Gabadadze} 
	B. Bajc and G. Gabadadze, Phys. Lett. {\bf B474} (2000) 282 
	[hep-th/9912232].
\bibitem{Daemi_Shaposhnikov} 
       	S. Randjbar-Daemi and M. Shaposhnikov, 
	Phys. Lett. {\bf B492} (2000) 361 [hep-th/0008079].
\bibitem{Dubovsky_Rubakov_Tinyakov} 
	S. L. Dubovsky, V. A. Rubakov, and 
	P. G. Tinyakov, Phys. Rev. D {\bf 62} (2000) 105011
	[hep-th/0006046].
\bibitem{Kehagias_Tamvakis} 
	A. Kehagias and K. Tamvakis, Phys. Lett. 
	{\bf B504} (2001) 38 [hep-th/0010112].
\bibitem{Ringeval_Peter_Uzan}
	C. Ringeval, P. Peter, and J.-P. Uzan, 
	Phys. Rev. D {\bf 65} (2002) 044016 [hep-th/0109194].
\bibitem{Koley_Kar}
	R. Koley and S. Kar, Class. Quantum Grav. {\bf 22} (2005) 753
	[hep-th/0407158].
\bibitem{Melfo_Pantoja_Tempo}
	A. Melfo, N. Pantoja, and J.D. Tempo,
	Phys. Rev. D {\bf 73} (2006) 044033 [hep-th/0601161].
\bibitem{Horava_Witten}
	P. Horava and E. Witten,
	Nucl. Phys.  {\bf B460}, 506 (1996) [hep-th/9510209];
	Nucl. Phys.  {\bf B475}, 94 (1996) [hep-th/9603142]
\bibitem{Lukas-Ovrut-Waldram}
	A. Lukas, B.A. Ovrut, K.S. Stelle, D. Waldram
	Phys. Rev. D{ bf 59}, 086001 (1999) [hep-th/9803235];\\
	A. Lukas, B. A. Ovrut, and D. Waldram,
	Phys. Rev. D {\bf 60} 086001 (1999) [hep-th/9806022].
\bibitem{Khoury_Ovrut_Steinhardt_Turok}
	J.~Khoury, B. A.~Ovrut, P. J.~Steinhardt and N.~Turok,
	Phys.\ Rev.\ D {\bf 64}, 123522 (2001) [hep-th/0103239]; \\
	J.~Khoury, B. A.~Ovrut, N.~Seiberg, P. J.~Steinhardt and N.~Turok,
	{\it ibid}.\ D {\bf 65}, 086007 (2002) [hep-th/0108187]; \\
	J.~Khoury, B.~A.~Ovrut, P.~J.~Steinhardt and N.~Turok,
	{\it ibid}.\ D {\bf 66}, 046005 (2002) [hep-th/0109050]; \\
	A.~J.~Tolley and N.~Turok,
	{\it ibid}.\ D {\bf 66}, 106005 (2002) [hep-th/0204091]. 
\bibitem{Steinhardt_Turok}
	P.~J.~Steinhardt and N.~Turok,
	Phys.\ Rev.\ D {\bf 65}, 126003 (2002) [hep-th/0111098]; \\
	J.~Khoury, P.~J.~Steinhardt and N.~Turok,
	Phys.\ Rev.\ Lett.\  {\bf 92}, 031302 (2004) [hep-th/0307132].
\bibitem{Anninos_Oliveira_Matzner}
	P.~Anninos, S.~Oliveira and R. A.~Matzner,
	Phys.\ Rev.\ D {\bf 44}, 1147 (1991).
\bibitem{Takamizu_Maeda1}
	Y.~Takamizu and K.~Maeda, Phys. Rev. D {\bf 70} (2004)  123514
	[hep-th/0406235]. 
\bibitem{Copeland}
	N.~D.~Antunes, E. J.~Copeland, M.~Hindmarsh and A.~Lukas,
	Phys.\ Rev.\ D {\bf 68}, 066005 (2003) [hep-th/0208219]; \\
	N. D.~Antunes, E. J.~Copeland, M.~Hindmarsh and A.~Lukas,
	{\it ibid}.\ D {\bf 69}, 065016 (2004) [hep-th/0310103].
\bibitem{Takamizu_Maeda2}
  	Y.~Takamizu and K.~Maeda,  Phys.\ Rev.\ D {\bf 73} (2006)  103508
	[hep-th/0603076].
\bibitem{Langlois_Maeda_Wands}
	D. Langlois, K. Maeda, D. Wands, Phys. Rev. Lett.{\bf 88},  181301 
 	(2002) [gr-qc/0111013].
\bibitem{Gibbons_Lu_Pope}
	G.W. Gibbons, H. Lu, C.N. Pope,
	Phys. Rev. Lett. {\bf 94}, 131602 (2005)
	[hep-th/0501117].
\bibitem{Chen_Chong_Gibbons_Lu_Pope}
	W. Chen, Z.-W. Chong, G.W. Gibbons, H. Lu, C.N. Pope, 
	Nucl. Phys. {\bf B732}, 118 (2006) [hep-th/0502077].
\bibitem{McFadden_Turok_Steinhardt}
	P.L. McFadden, N. Turok, P.J. Steinhardt, preprint [hep-th/0512123]. 
\bibitem{eto_sakai} 
	M.~Eto and N.~Sakai,
	Phys.\ Rev.\ D {\bf 68}, 125001 (2003) [hep-th/0307276]. 
\end{thebibliography}
\end{document}